\documentstyle[aps,twocolumn,psfig]{revtex}
 
\title{Object orientation and visualization of physics in two dimensions}
\author{Mark Burgess,H\aa rek Haugerud and Are Strandlie}
\address{Centre of Science and Technology, Faculty of Engineering, Oslo College, 0254 Oslo, Norway\\and\\Institute of Physics, University of Oslo, P.O.Box 1048 Blindern, 0316 Oslo, Norway}

\date{\today}

\newcommand{\abs}[1]{\mid\!\!#1\!\!\mid}
\newcommand{\be}{\begin{eqnarray}}
\newcommand{\beq}{\begin{eqnarray*}}

\newcommand{\ee}{\end{eqnarray}}
\newcommand{\eeq}{\end{eqnarray*}}

\newcommand{\ra}{\rightarrow}

\begin{document}
\maketitle

\begin{abstract}
We present a generalized framework for cellular/lattice based
visualizations in two dimensions based on state of the art computing
abstractions. Our implementation takes the form of a library of
reusable functions written in \verb-C++- which hides complex graphical
programming issues from the user and mimics the algebraic structure of
physics at the Hamiltonian level. Our toolkit is not just a graphics library
but an object analysis of physical systems which disentangles separate
concepts in a faithful analytical way. It could be rewritten in other
languages such as Java and extended to three dimensional systems
straightforwardly.  We illustrate the usefulness of our analysis with
implementations of spin-films (the two-dimensional $XY$ model with and
without an external magnetic field) and a model for diffusion through
a triangular lattice.
\end{abstract}

\section{Introduction}

Although computer simulations of many physical systems are common for
computing specific quantities, they do not necessarily give a direct
and overall impression of the dynamics of the systems under a variety
of conditions.  What is often lacking from the physicist's repertoire
is the possibility of a direct visualization of the behaviour of
microscopic systems in some appropriate semi-classical limit. Such a
mental image of the mechanics of the problem can be a source of great
inspiration, both for understanding established problems, and for
designing new scenarios. The ability to vary initial and boundary
conditions (temperature, external field and any other external
parameters) and compare several simulations has enormous potential for
the comprehension of the qualitative behaviour, while at the same time
enabling quantitative information to be calculated and displayed along
side.

In recent years the status of two dimensional physics has
changed from being mathematical idealization to being a realistic
physical prospect. The fractional quantum Hall effect and high
temperature superconductivity are widely believed to be
two-dimensional phenomena to a good approximation. Moreover
developments in the growth of ultra-thin layered heterostructures
makes the modeling of two-dimensional systems ever more important.

Theoretical developments in two dimensional physics have generated a
large body of work. Notable topics include the notion of fractional
statistics\cite{wilczek1} or anyons (particles which interpolate
between Bose-Einstein and Fermi-Dirac statistics), anyon
superconductivity and layered systems as potential models for the new
high temperature superconductors, the quantum Hall effect and spin
textures, to name but a few\cite{karlhede1}. Many of these models
could profitably be simulated visually to gain an intuitive picture of
what takes place. There are many other interesting candidates for
visual models in two dimensions: including the study of fields in
waveguides and cavities (the micromaser) and spin
computers\cite{feynman,zeigler,M.Minsky,zuse,schads,dahlberg1,argus,vlab}.
Spin diodes\cite{spindiodes} have been studied both experimentally and
theoretically and could profitably be made into a visual simulation.

Recently several groups have begun to appreciate the importance of
direct visualization in 2-dimensional systems.  Since direct
experimental observations are difficult to achieve, and certainly
difficult to repeat under identical conditions, computer simulations
are an obvious and valuable surrogate. Computer visualization always
involves some form of compromise or approximation, and sufficient
control of these compromises is an important concern, but the bonus of
a concrete dynamical picture of a physical system often outweighs
minor qualms about their accuracy. Graphical representation of data
places heuristic understanding before precision.

Certain physical situations are ideally suited to visualization.
For example: any kind of field, be it of a scalar or vector
character, with arbitrarily complicated boundary conditions, is easily
rendered visually, but might be described by complicated special
functions algebraically or numerically, making it difficult to gain a
qualitative understanding.

In this article, we present a visualization scheme for 2 dimensional
systems and layered 3 dimensional systems which may be used to play
physical simulations on a lattice like a one-way video
recording. Initial and boundary conditions may be edited, and parallel
simulations employing different parameters may be compared. The
framework is general, but we choose to illustrate it by looking at 2
dimensional spin systems and diffusion models, which provide a perfect
illustration of the principles.

\section{Cellular Automaton Simulation Environment (CASE)}

A convenient framework on which to base a system of visualization is
the cellular automaton.  The idea of cellular automata was introduced
around 70's and 80's with key papers by among others Wolfram. 
For a review, with references, see ref. \cite{wolframbook}.
A cellular automaton may
be thought of as a simulation engine in which cells, (or lattice
plaquettes) arranged in a symmetrical pattern, interact with their
neighbours according to well-defined rules. Each cell or site has a
number of variables associated with it, describing the state of the
system at that point (e.g. temperature, occupation number, spin state
etc), and that state evolves in time according to a supplied rule
which embodies the dynamics of the model. A suitable visual
representation of the automaton can be constructed by choosing a property like
colour to represent temperature (or other scalar quantities) and
arrows to represent vector quantities. Combinations of these, together
with the ability to limit and change the variable being displayed,
allow complex models to be visualized in detail.

Cellular automata have traditionally been used by physicists and
biologists for studying such diverse problems as the development of
cellular life, mixing of fluids, penetration of porous media, magnetic
spin systems etc. Any system in which space and time can be
discretized into an appropriate lattice can be modeled by a suitably
complicated automaton and this is in tune with the unavoidable
granulization which any system must undergo in a visualization.
Increasingly cellular automata are being used as simulated `analogue
computers' by engineers and statisticians. They are used in the
modeling of traffic flow at junctions, electrical networks, and the
diffusion of gases and spreading of interfaces, to name but a few
applications. It has been recently shown that there is a connection
between cellular automata and the $N$-soliton problem which is of
direct interest in non-linear optics as well as other
fields\cite{soliton}.  By adopting a generalized cellular automaton
for our simulation environment, we must introduce only one pertinent
restriction: namely that simulations are always pinned to a predefined
lattice\cite{footnote1}.

\section{Abstraction}

The most important technical feature of the CASE library is its object
orientated construction. Object orientation is about hiding details
inside `black-boxes' so that they do not interfere with the natural
logical structure of a problem. It is also about separating
independent issues in a program in a disciplined way.  Using an object
orientated philosophy together with the polymorphism allowed by C++,
we are able to create models with a structure which is independent of
low-level details. By low level details we refer to both microscopic
physical details of the models we simulate and low level programming
details, such as the specifics of how to draw graphics in windows. The
object model is a powerful abstraction which had largely been ignored
by physicists who are more used to nuts and bolts programming with
languages like FORTRAN. We feel that this aspect of CASE should not be
underestimated.  Practical models of real physics in complex systems
will only succeed in the future if there is a serious attempt to deal
with this complexity in a logical and organized manner.

As an example of the benefits which object orientation offers, we can
note the following. Normally cellular automata in two dimensions live
on a square lattice. In our framework we have been careful to admit
the possibility for models to employ any regular tesselating
lattice. We do this by separating issues which concern the lattice
structure from as much as possible of the remainder of the program,
and vice versa.  In many cases we can code physical models in a way which is
independent of the underlying symmetry of the lattice by referring only
to generic concepts such as nearest neighbours. The knowledge of how
to locate nearest neighbours can be hidden inside a black box
associated with a particular symmetry. This allows us to model
crystalline structures with hexagonal and other symmetry groups simply
by replacing one black box with another.
CASE is not simply a two dimensional graphics rendering package.  The
library is not just about making available different geometries.  It
is also about simplifying and rendering different physical
concepts. Any useful visualization technology must be based on
general, reusable abstractions which make physical understanding
paramount.  Our approach is based on a
fundamental untangling of concepts associated with two dimensional
cellular automata. 

Figure \ref{fig1} shows the way in which we have chosen to separate the
issues in the CASE library. The structure of objects and the way in
which they relate to one another have been designed with a practical
eye: we are looking to write real computer programs, not merely
theoretical constructs. The top of the hierarchy is therefore the
coarsest object in our scheme: an application, or complete program.
The application part of our scheme is a framework in which we can open
one or more simulations and run them concurrently, even in parallel.
An application contains convenient control switches for starting and
stopping simulations and keeping them synchronized with one another.

Beneath the application layer is the model layer. A model is a box
which encapsulates everything about a given physical system, such as
the rules by which physics is done. We can think of this part of the
model as defining the Hamiltonian for the system, since the rules of
how the system develops in time reside in this structure.  This black
box is itself built out of component objects which have a more general
validity than one single model. For instance, `Environment' describes
the symmetry of the lattice and the way in which one cell relates to
the others. This is analogous to the symmetry group of the lattice and
its boundary conditions (periodic, twisted, or isolated identification
of the edges).  A `Cell' object is a capsule which contains all of the
physical variables which are required by the physical model at every
site on the lattice. This is analogous to the choice of variables in
the Hamiltonian, or the parameterization of the problem. Finally we
have some strictly technical objects which have to do with the visual
mechanisms of the computer and the definition of re-usable symbols for
representing the physical variables contained in cells.

This object abstraction covers a complete classification of properties
associated with physical systems in a rational, formalized scheme.  We
can define plug-in visualization schemes for scalar, vector or other
data based on a lattice of arbitrary symmetry. We can render these
properties with arbitrary shapes and colours or even with numerical
values, or combinations of these objects. Visual objects might hide
variables which can be revealed by clicking on a cell to avoid visual
clutter.  Perhaps more importantly than the results, we are able to
program the physics in a way which preserves the structure of a system
at the algebraic level: cells are the basic dynamical variables,
models are the Hamiltonian, and each of these can be separated,
without muddling issues.

\vspace{0.5cm}
\begin{tabular}{|c|c|c|}
\hline
CASE & Symbol & Meaning\\\hline\hline
{\tt ~~Update()}~~& ~~$H$~~ & ~~Hamiltonian~~ \\\hline
{\tt Cell} object & $\phi$  &  The field \\\hline
{\tt Environ} & $\cal G$ & Symmetry group \\\hline
{\tt Property} & $T,E,{\bf S}$ & Physical quantities \\\hline
\end{tabular}
\vspace{0.5cm}

The basic symbol types which have been used so far are in example
models are: coloured blobs to represent scalar values, arrows (with
variable size) to represent vector data and the use of scaled and
coloured plus and minus symbols to represent the locations of positive
and negative charges of varying magnitudes, particularly in connection
with vortices or charges.

\section{Implementation}

The CASE simulation environment was began in Oslo in 1993\cite{demo}
with a simple implementation of the $XY$-model programmed as a one-off
application on a unix workstation, using the X11 windowing
library\cite{X11}.  Since then, we have developed a generalized
framework for cellular or lattice based visualizations, based on state
of the art, freely distributable computing
systems\cite{case,footnote2}.

Our aim has been to create a flexible system for constructing
generalized, visual models sufficient for our own special needs, but
which may be adopted and developed by others, without the need for
expensive software packages. Indeed all of the tools required to use
our models are freely available on the internet. Our code is written
in object-oriented C++ and is designed carefully to be comprehensible
and reusable by the physics community. The technical details of this
framework will be published elsewhere\cite{case}. CASE takes the form
of a library of reusable functions which hide complex programming
issues from the user. We envisage our user to be an intelligent
scientist with an aptitude for programming, but with little patience
for incomprehensible graphical window interfaces.  The framework may
be loosely described as follows:

\begin{itemize}

\item CASE supports an $N$ state cellular automaton, based
on a lattice with user configurable symmetry. Each site can be
edited or randomized to specify initial conditions.

\item The method of visualization must be user configurable. Simple
mechanisms for adding and changing colour in real-time are provided,
for example.

\item The size of system and choice of boundary conditions may be
changed as run-time parameters.

\item Resizing and zooming in and out of the lattice permits the
handling of large models.

\item Freeze frame and resume, with possibility of saving the state
of the simulation for later continuation or analysis. Snapshots may
also be sent to a printer or other device.

\end{itemize}

\subsection{Object construction}

The implementation of a model in CASE involves the definition
of a model object (which in turn uses reusable cell objects),
and all of its methods (functions). Here is an example
model object.

\small\begin{verbatim}
class XYModel : public CAModel
{
public:

   XYModel();
   XYModel(Widget new_parent);
   ~XYModel();

   virtual void Update();
   virtual void RightClick(double x, double y);
   
protected:
   
   virtual void Draw(int cell_index);
   virtual void Redraw();

   virtual void AllocateCells(int new_cells);
   virtual void ReadEnvironRequester();
   
   void Init_spinconfig();
   int Monte_Carlo(int cell_index);
   double Trial_angle(XYCell *current);
   double New_energy(int cell, double trial_angle);
   double Old_energy(int cell);
   int CheckVortex(int cell_index);
   void  CheckVortices();

   CAArrowSymbol *symbol;
   
   double beta;
   Req temp_req;
   int iter;
   
private:
   static void temp_ok(Req *req, XtPointer data);
   static void temp_cancel(Req *req, XtPointer data);
   static void set_temp(Widget w, XtPointer data,
                         XtPointer garb);

};

\end{verbatim}\normalsize
The function prototypes here refer to functions which the user of CASE
must supply in order to describe the nature of cells.  In addition to
certain required functions which control aspects of the user
interface, which CASE needs in order to function ({\tt RightClick},
{\tt Update}, etc.), any number of other functions may be created as
is convenient for implementing the model. In the $XY$ example, these
include {\tt Init\_spinconfig}, {\tt CheckVortices} etc.

The most important function is the Update method. This
is a formal coding of the Hamiltonian of the system.
In many cases this update procedure is based upon
randomized statistical processes. This is implemented
using a Monte Carlo algorithm.

\subsection{Monte Carlo update procedure}

In a statistical system, close to thermodynamic equilibrium, the
average state of the system is characterized by a temperature $T>0$,
or equivalently by inverse temperature $\beta$. Microscopically, the
state is defined by the condition of every spin in the simulation, but
this is not a static state: we must allow for fluctuations. Since it
is impractical to account for every degree of freedom which gives rise
to such fluctuations, one adopts a heuristic algorithm for
including them. In our case, this is known as the Metropolis
algorithm, which is a variant of so-called Monte Carlo methods.

A Monte Carlo method is a way of using random numbers to update the
state of the system (the name Monte Carlo evokes images of gambling).
Instead of iteratively parsing the cellular grid and updating state of
each cell deterministically, one chooses spins in the grid according
to some rule and applies a rule which only updates spins only with a
certain probability, characterized by the Boltzmann factor. The
algorithm is as follows:
\begin{itemize}
\item Select a spin,
\item Pick an angle by which the spin might flip at random,
\item Compute the energy change associated with the random flip $\Delta E$,
\item Calculate the probability of transition $W=\exp(-\beta\Delta E)$,
\item Calculate a random number $x$ between zero and unit,
\item if $x < W$, update the spin, otherwise leave it alone.
\end{itemize}

In a real system, fluctuations cause spins to flip at random. In this
sense the Monte Carlo algorithm may be thought of as a true
time-dependent simulation of a physical process approaching
equilibrium. Note however, that there is no immediate connection
between the time elapsed in the simulation and `physical time'. The
true physical time is modelled most realistically if we update the
sites at random, but time is saved if we go through the grid in an
orderly fashion. Although one would not expect any major problems to
arise from this orderly parsing of the grid, it makes the
identification of the physical time a more heuristic than precise
procedure. In the limit of large times, the ergodicity of the system
should guarantee that the specific method of updating would not play a
role in the final outcome. In our example implementation of the
XY-Model, we choose random updates to best simulate a physical system.

\subsection{Threads}

In the present model we have been
able to get away with a rather primitive algorithm for scheduling
updates of the window graphics, based on the X timeout mechanism.
This method has several failings: if the calculations are very time
consuming, then the normal X event loop becomes neglected.  This means
that the responsiveness of windows and button clicks with degrade and
become unacceptably slow. This is likely to be a problem in more
ambitious simulations. The only way to avoid this problem is to
separate the updating of the window from the business of processing
X11 messages. The most natural way to do this is to use threads.  By
making a simulation multithreaded, we can most efficiently organize
the time spent on each task without having to break up the updating
algorithm, in an unaesthetic fashion. This is also a natural objection
oriented solution to the problem.

CASE currently uses release 6 of X11 and will continue to build on
this and later releases. X11R6 contains a thread
compatible library which we shall experiment with in the future.  One
possibility is to use POSIX pthreads, but so far only a few operating
systems have implemented pthreads in an acceptable way. We shall
most likely try to encapsulate the threading abstraction in a suitable
class to hide the chiefly technical difficulties.

\section{Example visualizations}

We illustrate the usefulness of our simulator with examples based on
two-dimensional spin-films and a diffusion model.  We hope that these
model simulations will be of widespread interest to physicists and
students alike and lead to many helpful insights into the nature of
microscopic systems both in classical, semi-classical and quantum
mechanical descriptions. In the first two cases we focus on what
can be learned from the visualization of physical systems, while in
the final example we demonstrate the benefits of our abstraction
policy.

\subsection{The two-dimensional $XY$ model}

One of the simplest but nonetheless interesting models in two
dimensions is the so-called $XY$-model for spin systems. This model
has been studied many times before in simulations, but to our
knowledge never visualized directly. It exhibits vortices and a phase
transition and therefore serves as a basic template for all spin
models in two dimensions.  Considerable efforts have been expended in
order to explore the properties of the two-dimensional classical $XY$
model (for a review, see
\cite{XYREW}). This has partly been motivated by the fact that the
$XY$ model can be viewed as an approximate microscopic model of a
neutral two-dimensional superfluid\cite{SUPERF} and thus describe
physical systems such as $^4 He$ films.  Furthermore it is closely
related to the two-dimensional Coulomb gas, which is also capable of
describing charged superfluids\cite{MinnhRev}. When including the
electromagnetic vector potential in the $XY$ model it can be viewed as
an approximate microscopic model for superconducting films as well as
for two-dimensional arrays of weakly coupled Josephson junctions
\cite{LA80}.

Due to the complexity of the models, analytic solutions are
hard to find; one therefore resorts to Monte Carlo simulations
to obtain estimates of thermal expectation values
\cite{Tobochnik,Schultka,BINDER92}.

The two-dimensional $XY$ model is a model of classical spins $
{\bf S} $ located at each lattice site of a two-dimensional square
lattice with lattice spacing $ a $.  The name $XY$ stems from the
fact that the spins are constrained to rotate in a plane, and the
spins are classical in the way that they can point in any direction in
this plane. Originally the model was derived as a simplification of
the Heisenberg spin model of ferromagnetism.  In the classical
Heisenberg model the spins can point in any direction in
three-dimensional space.
  
The basic feature of the model on a microscopic scale is the nearest-
neighbour interaction between the spins. The Hamiltonian is given by
\begin{equation}
   H= -J \sum_{<{\bf x,x'}>} {\bf S_{x} \cdot S_{x'}}
   \stackrel{|{\bf S}| =1}{=} -J \sum_{<{\bf x,x'}>} 
   \cos ( \phi_{{\bf x}}- \phi_{{\bf x'}} ) 
\end{equation}
where $<{\bf x,x'}>$ means that the sum is restricted to
nearest-neighbours, and $ J $ is a positive coupling constant making
the coupling ferromagnetic. Here $ \phi_{{\bf x}} $ means the angle $
{\bf S_{x}} $ makes with an arbitrary, fixed axis.  The line between
two sites will be denoted a link, and the square region on the inside
of four neighbouring links will be denoted a plaquette. An
illustration of such a spin lattice is shown in figure \ref{fig2}.

In the CASE setup, this Hamiltonian is coded into a module called
XYModel which makes reference to separate cell objects containing
angle data.  The statistical mechanics of the model is contained in
the partition function
\begin{equation}
   Z= \int_{-\pi}^{\pi} \prod_{{\bf x}} d\phi_{{\bf x}}
   e^{K \sum_{<{\bf x,x'}>}
   \cos ( \phi_{{\bf x}}- \phi_{{\bf x'}} ) } 
\end{equation}
where $ K = \beta J = J/kT $.
Any thermodynamic quantity of interest can be calculated through the 
partition function, or in our case from the Model abstraction.

We now summarize some of the physics of this model.
If we transform each spin according to
\begin{equation}
   \phi_{{\bf x}} \rightarrow \phi_{{\bf x}} + \alpha 
\end{equation}
where $ \alpha $ is some fixed angle, we see that the Hamiltonian is
invariant. The system possesses therefore a continuous symmetry. When
$ T \rightarrow 0 $ all spins will be aligned because of the
ferromagnetic interaction, and hence the model has an infinitely
degenerate ground state.  The ground state configuration is where all
spins point in the same direction, thus we expect only small
deviations from such a configuration at low temperatures. This means
we may write
\begin{equation}
   \phi_{{\bf x}}- \phi_{{\bf x'}} \ll 1,
\end{equation}
where $ {\bf x} $ and $ {\bf x'} $ are neighbouring sites. Furthermore, it
suggests that we can expand the cosine in a Taylor series and keep the
terms up to second order,
\begin{equation}
   \cos ( \phi_{{\bf x}}- \phi_{{\bf x'}} ) \simeq 1- \frac{1}{2} 
   ( \phi_{{\bf x}}- \phi_{{\bf x'}} )^{2} 
\end{equation}
and thus
\begin{equation}
   \label{XYHamiltonian1}
   H= \frac{J}{2} \sum_{<{\bf x,x'}>} ( \phi_{{\bf x}}- 
   \phi_{{\bf x'}} )^{2} 
\end{equation}
Here an unimportant constant term expressing the zero point energy has been discarded.
  
If we look at the spin system on a large scale, it is not possible to
see the microscopic structure. We can therefore approximate the discrete 
formulation with a continuum model. This is equivalent to letting the
lattice spacing approach zero, and hence
\begin{equation}
   H= \frac{J}{2} \sum_{<{\bf x,x'}>} ( \phi_{{\bf x}}- 
   \phi_{{\bf x'}} )^{2} \stackrel{a \rightarrow 0}{\rightarrow} 
   \frac{J}{2} \int d^{2}x 
   (\nabla_{\mu} \phi)^{2}
\end{equation}
where $ \mu=1,2 $.  The Hamiltonian is seen to have the same form as
the action of a real, one-component, massless field in 2+0
dimensions. The partition function then becomes the functional integral
\begin{equation}
    Z= \int {\cal D}\phi e^{-\frac{K}{2} 
    \int d^{2}x (\nabla_{\mu} \phi)^{2}} 
\end{equation}
This changeover from a discrete to continuum model can be simulated in CASE
by turning up the number of sites on the lattice. We are also able to
zoom in and out of the lattice to compare and contrast the microscopic and
the macroscopic.
  
The field configurations which make the action stationary with respect
to small variations of the field can be found as solutions of the
Euler-Lagrange equations
\begin{equation}
   \frac{ \delta H[\phi]}{ \delta \phi} =0
\end{equation}
which implies
\begin{equation}
   \nabla^{2} \phi=0 
\end{equation}
By transforming to polar coordinates, two simple solutions are easily guessed
\begin{eqnarray*}
  1) & ~\phi & = const \\
  2) & ~\phi & = n \cdot \theta
\end{eqnarray*}
Here $ \theta $ is the polar angle with respect to some origin.  The
first solution gives us the global energy minimum of the system with
all spins aligned. For the second solution there is mathematically no
constriction on the real constant $ n $, but the fact that the field
strictly is an angle gives
\begin{eqnarray}
     \phi(\theta + 2\pi) = \phi(\theta) + k \cdot 2\pi, & k \in
     {\cal Z}
\end{eqnarray}
and hence $ n=k $.
If $ n $ is chosen to be $ 1 $, a spin configuration will look like the one
shown in figure~\ref{Vortex}a.
This configuration is what is called a vortex. By letting each spin
transform according to $ \phi \rightarrow \phi +
\pi/2 $ we still get a solution of the equation, and a typical
whirl shows up. This is shown in figure~\ref{Vortex}b. We expect
to find such vortices in a simulation.
  
We see that, in moving along the whole boundary of the lattice in
figure~\ref{Vortex}a counterclockwise, the spins have undergone one full
revolution. If the sites are labeled $ (x,y) $ with $ (1,1) $ at
the bottom left and $ (4,4) $ at the top right, the sum of spin
differences along the boundary is given as
\begin{eqnarray}
   \label{SumSpinDifferences}
   S &=& (\phi_{2,1}-\phi_{1,1}) + (\phi_{3,1}-\phi_{2,1}) \nonumber\\
   &+& (\phi_{4,1}-\phi_{3,1}) + ....... + (\phi_{1,1}-\phi_{1,2}) = 2\pi
\end{eqnarray}
with $ S = 2\pi $ expressing the fact that the spins have undergone 
one revolution. In
general, evaluating this sum of spin differences along any closed path on a
lattice will give some multiple of $ 2\pi $
\begin{equation}
   S = n \cdot 2\pi
\end{equation}
and this $ n $ is the definition of the vorticity or the charge of the
vortex. If $ n=0 $ there is no vortex inside the loop. If $ n \neq 0
$, there is a vortex with vorticity $ n $ inside the loop. When there
are several vortices inside our closed loop, the value of $ n $
becomes the total vorticity of the region. The vortex in
figure~\ref{Vortex}a is seen to have $ n = 1 $, and it follows that a
vortex with vorticity $ n $ is constructed in the simplest way by
letting
\begin{equation}
   \phi = n \cdot \theta
\end{equation}
  
According to Kosterlitz and Thouless\cite{KT73} the vortices only
exist in tightly bound pairs with opposite vorticity at low
temperatures. At some critical temperature, $ T_{c} $, the first
vortex pair unbinds and the vortices are free to move to the
one-dimensional surface of the lattice under the influence of an
arbitrarily weak, external magnetic field. The phase transition is
therefore called the vortex-unbinding transition, and it is
characterized by this abrupt change in the response to the magnetic
field. Following their argument, the vortices are equivalent to freely
moving, straight parallel, current-carrying conductors located at the
centres of the vortices. These will therefore tend to move at straight
angles to an applied magnetic field in the plane of the system, the
direction determined by the direction of the current. This is easily
seen by the well-known formula for the force on a current-carrying
wire in a magnetic field
\begin{equation}
   {\bf F} = I \cdot {\bf l} \times {\bf B}
\end{equation}
which tells us that the forces on wires with oppositely directed currents 
are oppositely directed and perpendicular to the field.
In the case of a vanishing external magnetic field there are no driving
forces on the vortices, but we should still observe free vortices above 
$ T_{c} $ solely due to thermal fluctuations.
 
In order to focus on the thermal behaviour of the XY model we use a
Monte Carlo simulation, with a standard, local Metropolis
\cite{Metropolis} updating scheme which chooses new trial
configurations by a random change in the direction of one single spin
at the time. Our simulation has provided a new {\em visual} way of
examining the physics in the $XY$ model, and it gave us some unknown
and interesting information about the system's behaviour in the
non-equilibrium phase of the Monte Carlo simulations. When starting
from an initial, random spin configuration, it appeared at low
temperatures that the system always went through a vortex phase before
settling in an equilibrium configuration, where the spins pointed all
in very nearly the same direction. We also could see that the vortices
disappeared only by drifting into others with opposite vorticity. This
behaviour was not observed at higher temperatures.
  
Figure \ref{XYmodel1} shows snapshots of the terminal screen during a
low temperature Monte Carlo simulation.  The system is seen to start
out from an initial, random configuration. Soon the spin configuration
is dominated by the vortices, and they start annihilating by drifting
into one another.  Figure \ref{XYmodel1}b shows a configuration, close
to equilibrium which contains few vortices.  These persist due to the
nature of the Metropolis Monte Carlo algorithm and also a consequence
of the periodic boundary conditions. The Metropolis algorithm only
checks the couplings to the nearest neighbours when trying to update a
spin, and it is therefore a local updating algorithm. When we are at
very low temperatures only configurations with lower energy will be
accepted. This means that applying the algorithm can be thought of as
combing a disordered, long-hair carpet. Because Metropolis only combs
locally, it will start a local ordering of the spins all over the
system, creating small domains of aligned spins.  The ordering will
gradually grow to bigger regions, and the vortices will show up. In
the end all of the spins will have to be aligned at these low
temperatures as in figure \ref{XYmodel1}c, and the only way to get rid
of the vortices is by combing oppositely charged pairs into one another.
  
In addition the vortex phase turned out to be very stable. It only
took a few iterations to establish this phase, but many more were
required to ``comb out'' the vortices into the smooth equilibrium spin
configuration.  The stability can be explained by the fact that the
vortices appear as configurations having stationary energy with
respect to variations of the field
\begin{equation}
   \frac{\delta H[\phi]}{\delta \phi}=0 
\end{equation}
This means that the vortices become local potential wells, and by
trying to get out of such a configuration one will almost always have
to increase the energy of the system. It is then reasonable that we
will need a considerable number of Monte Carlo steps to take the
system out of these local energy minimum configurations into the
equilibrium, global energy minimum. 
  
We were also able to directly probe the Kosterlitz-Thouless prediction
of the model having two different phases in equilibrium using the
visual simulation.  It appeared that the vortices exist only in
tightly bound pairs below the critical temperature. As expected, we
were able to detect free vortices above the critical temperature. Just
above $ T_{c} $ the event of a vortex pair unbinding was very rare,
but the unbinding clearly became more common as the temperature was
increased further.
  
Figure~\ref{Video4} shows some equilibrium snapshots above and below $
T_{c} $.  The spins are here omitted in order to emphasize the vortex
behaviour.  In figure \ref{Video4}a we display a typical
configuration below the critical temperature. Vortex pairs are
spontaneously created and annihilated in pairs, and they always exist
in pairs. For the case of a temperature above the critical temperature
a different behaviour is observed, as shown in \ref{Video4}b.  
The vortices are also here created and annihilated in pairs, but
we have here the possibility of a vortex pair unbinding.  
This event is very rare just above $ T_{c} $, but it is more
common at higher temperatures. The effect seems to be caused by other
vortex pairs being created in between the original pair. These other
pairs are then polarized and thus screen off the interaction between
the original pair, thereby making it easier for the original pair to
unbind. 

\subsection{The two-dimensional $XY$ model in a magnetic field}

In this section we present the simulation of a model which is quantum
mechanical in nature. To our knowledge there have been no previous
attempts to construct a visual simulation of this model.

Superconductors exhibit the Meissner effect. In the
superconducting phase an external magnetic field $H$ will be repelled
and within the superconductor the magnetic field is zero. In a type-I
superconductor, upon increasing the external magnetic field, the
superconducting state collapses at the critical field $H_c$ and the
applied field enters the material.  But if the external field is now
reduced the material re-enters the superconducting state at $H_c$, and
the flux which had entered is expelled. This expulsion of the flux,
the Meissner effect, would not occur for a perfect conductor.  In a
type-II superconductor flux first enters the superconductor in the
form of quantized flux lines at a critical field $H_{c1}$, but the
superconducting state does not collapse.  As the applied field
increases the density of flux lines increases until the system enters
the normal state smoothly at the upper critical field $H_{c2}$.  A
$H-T$ phase diagram for a type-II superconductor is sketched in
figure. \ref{H-T}.

The phenomenological Ginzburg-Landau theory of superconductivity has
very successfully explained all the most important features of
superconductivity \cite{GL}.  It's basic assumption is that a
superconductor at each point in space is characterized by a complex
order parameter $\psi({\bf x})$
\be
\psi({\bf x}) = \abs{\psi({\bf x})} e^{i\phi(x)}
\ee
where $\abs{\psi({\bf x})}^{2}$ equals the density of superconducting
Cooper pairs.  There are two characteristic lengths in the
Ginzburg-Landau theory.  A magnetic field will penetrate a small
distance into the superconductor, but vanishes at a distance $\sim
\lambda$; the penetration depth.  The coherence length $\xi$ is
roughly the distance over which the superconducting phase $\psi({\bf
x})$ varies from zero to it's maximal value.  A superconductor can be
characterized by the Ginzburg-Landau parameter $\kappa = \lambda/
\xi$.  If $\kappa < 1/\sqrt{2}$ it is type-I, otherwise it is type-II.
The high-$T_c$ superconductors are of extreme type-II in the sense
that the $\kappa \gg 1$.  It was shown in a classic paper by Abrikosov
\cite{ABRI57} that the flux lines in a type-II superconductor will
form a hexagonal lattice and this has also have been seen
experimentally \cite{ABRIEXP}. The dynamics of such a 3 dimensional
flux-line or vortex lattice has been of considerable interest after
the discovery of high temperature superconductors \cite{REVIEWS}.
Based on experimental \cite{EXP} and theoretical
\cite{THEORY,HPS89,2Da} results it was proposed that the flux-line
lattice melts into a vortex liquid over large parts of the $H-T$ phase
diagram sketched in figure \ref{H-T}.

A widely used microscopic model \cite{MODEL,ASLE,FRANZ95} for
describing the statistical mechanics of the flux line lattice is the
$XY$ model in a magnetic field with a Hamiltonian given by
\be
H = -\sum_{ij}\cos(\phi_i - \phi_j - A_{ij})
\ee
where $\phi_i$ is the phase of the superconducting wave function $\psi$ at site $i$, the sum 
is over nearest neighbour sites and 
\be
A_{ij} = \frac{2e}{\hbar c}\int_{i}^{j} {\bf A}\cdot d{\bf l}
\ee
is the integral of the vector potential from site $i$ to site
$j$. This effective model can be derived by discretizing the
Landau-Ginzburg free energy functional under the assumption of a
spatial uniform magnetic induction and constant $\abs{\psi}$ outside
the normal vortex cores, the London approximation \cite{LT93}.  The
flux lines are three dimensional objects, but in some cases it turns
out to be a good approximation to treat them as two dimensional,
i. e. as straight lines in the $z$ direction. A superconducting thin
film is one case and another is high-$T_c$ superconductors which
consist of weakly coupled layers, and so for some range of parameters
may display effectively two dimensional behaviour \cite{2Da,2Db}.  In
an extreme type-II superconductor the magnetic field, given by $B =
\nabla\times A$, is to a good approximation uniform, when the magnetic
penetration depth $\lambda$ is larger than the inter vortex distance.
In the model the fluctuations in the amplitude $\psi({\bf x})$ is also
neglected. In order to justify this, the inter vortex distance $a_0
\approx \sqrt{\Phi_0/B}$ must be much larger than the coherence
length. In the Ginzburg-Landau theory, the lower and upper critical
fields are given by $H_{c1} \approx \Phi_0/4\pi\lambda^2$ and $H_{c2}
\approx \Phi_0/2\pi\xi^2$. For an extreme type-II superconductors
$H_{c2}/H_{c1} \approx 2\xi^2 \gg 1$ and there is thus a wide field of
the $H-T$ phase diagram of figure ~\ref{H-T} for which the amplitude
fluctuations can be neglected \cite{SB91}.

We consider a quadratic lattice with lattice constant $a$ which serves
as a measure of the coherence length $\xi$. Each flux line carries a
flux equal the flux quantum $\Phi_0 = hc/2e$ and the average density
of field induced vortices is thus $f = Ba^2/\Phi_0$. Since the total
flux through a plaquette of the lattice equals $\oint {\bf A}\cdot
d{\bf l}$, the sum over $A_{ij}$ around such a square must obey the
constraint
\be
A_{ij} + A_{jk} + A_{kl} + A_{li} = 2\pi f
\ee 
The localization of the vortices are as in the standard $XY$ model
determined by calculating the total change of the phase $\phi$ around
a plaquette: $\sum_{\Box} (\phi_i - \phi_j) = 2\pi n$, where $n$ is
the integer vorticity. Neither the phase $\phi$ nor the vector
potential ${\bf A}$ are directly observable quantities, but choosing a
particular gauge fixes ${\bf A}$ and thereby $\phi$.  However, it is
convenient \cite{ASLE} to treat the model in a gauge invariant manner,
and we will follow this approach. The superconducting current density
is a physical observable and can be expressed as
\be
{\bf J} = \abs{\psi({\bf x})}^{2}\frac{e\hbar}{m} \left(\nabla \phi - \frac{2e}{\hbar c} {\bf A}\right)
\ee
Hence the factor $\alpha_{ij} = \phi_i - \phi_j - A_{ij}$ of the
Hamiltonian is gauge invariant. Throughout the simulations no
reference is ever made to a specific vector potential, nor to a
specific phase $\phi$.  The Monte Carlo moves are instead performed on
the gauge invariant phase $\alpha$ which is the single dynamic
variable of the system. As stated above, the vorticity depends on the
phase difference of $\phi$ only.  Nevertheless it is possible to
calculate the vorticity from the gauge invariant phase $\alpha$ since
\be
\sum_{\Box} (\phi_i - \phi_j - A_{ij}) = 2\pi( n - f) \label{LINKS}
\ee
where the value of $\alpha_{ij} = \phi_i - \phi_j - A_{ij}$ is
restricted to the interval $(\pi,-\pi)$.

One of the goals of Monte Carlo simulations of vortex lattices is to
the determine the phase diagram (see for instance figure 1 in
\cite{HW94}) including the pinned solid Abrikosov phase and a liquid
phase as well as the depinned floating solid phase.  For the latter
the hexagonal lattice is intact, but it is not pinned and able to move
as a whole.  In the simulations this corresponds to a phase where the
hexagonal lattice floats on the underlying numerical lattice.  For
such investigations it is very useful to be able to see how the actual
configurations change during the simulations, for different
temperatures and concentrations of vortices.

Here we will concentrate on the formation of a hexagonal Abrikosov
lattice generated when starting from high temperature and relaxing at
a sufficiently low temperature. For studies of the phase diagram it is
essential that a hexagonal lattice can be formed at low temperatures
without frustration.  In actual simulations the size of the lattices
is restricted by practical concerns and in order to approximate a
large system better, periodic boundary conditions are imposed.  If the
appropriate size of the underlying numerical lattice is not carefully
chosen, this will lead to frustration of an hexagonal lattice. Since
we use a quadratic numerical lattice, a perfect hexagonal lattice can
not be generated, and also a proper vortex concentration $f$ must be
chosen.  The dilute case, $f \ra 0$, can equivalently be viewed as the
continuum limit, in which the lattice spacing $a$ decreases to zero
for a fixed areal density of vortices. So for addressing the problem
of vortex lattice melting in a uniform superconducting film, the
vortex concentration $f$ should be small, probably less than $1/30$ in
order to see all the three phases of the phase diagram.  The ground
state configurations are quadratic for large concentrations
\cite{fBIG}, even for $f = 1/25$ \cite{f25}, and hence $f$ must be
smaller in order to approach the continuum limit.

These considerations are taken into account when choosing $f = 1/30$
on a lattice of $30 \times 30$ lattice sites \cite{f30}. For these
parameters an almost perfect hexagonal lattice may be formed without
frustration due to periodic boundary conditions.  As an initial state
we choose the high temperature limit, a random configuration of
vortices. It is possible to construct
an algorithm which initiates the lattice with any given configuration
of vortices by going sequentially through the lattice and assigning
phases on the links according to Eq. ~\ref{LINKS}, choosing $n = 1$
for a plaquette containing a vortex and $n = 0$ for plaquette
containing none. After the initial configuration has been loaded,
sites are chosen at random and probed by the Metropolis method. A
trial configuration for a single Monte Carlo step is chosen by adding
a random phase $\Delta$ to the links surrounding a site as shown in
 figure \ref{MCSTEP}.  A sequence of steps equal to the total number of
lattice sites, $30 \times 30$, is one iteration. During the first few
iterations the vortex configuration changes rapidly and occasionally
vortex anti-vortex pairs are created, as seen in
figure \ref{XYB}a. Even vortices with a vorticity of two may occur
in this chaotic phase of the simulation.  After a large number of iterations
an almost hexagonal lattice is formed, as seen in
figure \ref{XYB}b. There are some fluctuations around this
configuration because of the finite temperature, but the hexagonal
lattice is pinned to the underlying numerical lattice and thus not
moving. According to Ref. \cite{f30} one would expect the vortex
lattice to melt into a vortex liquid when increasing the temperature
above $T = 0.045$. At this temperature both the helicity modulus, a
measure of long-range phase coherence, and the sixfold orientational
order parameter drops rapidly to zero, which means that depinning and
melting occurs at the same temperature.

However, if the vortex concentration is lowered to $f = 1/56$ the
helicity modulus drops at $T_p = 0.03$ and the orientational order
parameter at $T_m = 0.05$ \cite{f30}. This means that the system at
$T_p$ enters the floating solid state where the hexagonal vortex
lattice is unpinned and moves as a whole independent of the numerical
square lattice. The appearance of such a phase signals the onset of
the continuum limit, since the pinned phase is an artifact of the
lattice model. This is because there is a finite energy cost for
deplacing a vortex in the ground state configuration.  At $T_m$ the
hexagonal lattice melts and the system finally enters the vortex
liquid phase. A vortex concentration $f$ of less than $1/30$ is
therefore necessary in order to study the true melting of the flux
lattice.

The gauge invariant phases $\alpha$ are also visualized in figure
\ref{XYB}.  The arrows show the positive direction of the phase
(current) and their size is proportional to the magnitude of the
phases. Close to the vortex core there are strong currents due to the
rapid change of the superconducting phase $\phi$. In between the
vortices the current around a plaquette is small, but nonzero. Here
there is no contribution from the superconducting phase, but there is
a small net current in the opposite direction due to the constant
magnetic field.

\subsection{Diffusion into a lattice}

In the previous examples the models were based exclusively on a square
lattice in order to simplify special highlighting algorithms. In this
section we wish to present a more generic automaton which demonstrates
the adaptability of CASE's object oriented construction. We use a toy
model for classical Brownian motion. The algorithm is intended to
represent diffusion of particles into the surface plane of a
crystalline lattice from outside.  The plane being visualized should
therefore be thought of as a surface layer separating a three
dimensional region of solid crystal from a corresponding region of
empty space filled with gaseous particles. When particles penetrate
the lattice from the gaseous region they appear spontaneously in the
model at a random site. When they leave lattice by further penetration
into the bulk or by escaping back into the gaseous region they
disappear from the model.  This is used as a simple toy model which
may be developed into a model for the way in which atomic hydrogen
penetrates a palladium lattice.

The geometry of the lattice is completely hidden from the model
algorithms and we can therefore run several simulations with different
lattice structures side by side and compare them without having to
rewrite any code whatsoever.  It is not necessary to know the
specifics about the background geometry: simple abstractions such as
`nearest neighbour' suffice.  To determine possible destinations for
particle hopping, we need only obtain a list of nearest neighbours
from the lattice-specific object. Figure \ref{diff}a illustrates the
diffusion model implemented on a rectangular lattice, while figure
\ref{diff}b shows the same model on a triangular lattice. The difference
between these images is a single switch: no special code is required
in the model to make this change. Figure
\ref{diff}c shows a larger simulation with some background colour
highlighting.

The dynamics of the model are derived from simple energy
considerations.  To make the model physically reasonable, we must have
an exchange of energy and a Monte-Carlo based algorithm for randomly
perturbing the system. For this simple demonstration we choose to
illustrate an interaction between the thermodynamical properties of
the lattice and the arrival and disappearance of particles. When a new
particle enters the system, it contributes an amount of energy which
warms up the lattice locally. This energy input is a combination of
kinetic energy and energy of dissociation of a chemical bond as
molecular hydrogen becomes atomized. The values of these energies are
set to constant values in the model object.  The extra heat
contributed to the lattice is visualized by a change in the background
colour from blue (cold) to green (warm). The change occurs locally,
but on each iteration of the model, an averaging procedure is used to
conduct away local hot spots and maintain thermal equilibrium. The
colour of the particle blobs represents the energy of the particles
and uses a different colour scale for clarity. The energy of the
particles is chosen randomly for the sake of illustrating a variety of
colours.

Particles are selected at random and can hop over to nearest neighbour
sites with a certain probability; this introduces a Brownian motion
into the system. The probability for hopping depends on an energy
barrier or work function which we have chosen to scale according to
the square root of the average temperature between nearest neighbour
sites. This crudely simulates a classical harmonic barrier of hight $x$
where $\frac{1}{2}k x^2 = k_BT$. Thus increased thermal activity of
the lattice makes diffusion harder at constant particle energy.
The following code snippet illustrates a simple Brownian motion
algorithm using the lattice independent mechanisms.

\small
\begin{verbatim}

 // Find the nearest neighbours at 1-st level
 // by querying lattice

 grid->QueryNeighbours(i, 1, neighbours);

 for (i = 0; i < neighbours.GetSize(); i++)
    {
    if (neighbours.array[i] == environ->NoCell)
       {
       continue; // Edge effect
       }
 
    neigh = (ThermoCell *) cell[neighbours.array[i]];
    
    if (neigh->Alive())
       {
       continue;
       }

    // Get energy for required processes
 
    beta = Tscale / (current->T + neigh->T);
 
    transition_prob = 
       exp(-beta*Work(neigh,current,cell_index,i));
 
    // Note `least action'

    if (transition_prob > highest)
       {
       highest = transition_prob;
       favourite = neighbours.array[i];
       }
    }
 
 // If we have no free neighbour sites
 
 if (highest == 0.0)
    {
    return;
    }
 
 // Move if the roulette wheel favours the wicked ...
 
 if (highest >= drand48())
    {
    neigh = (ThermoCell *) cell[favourite];
    current->SetAlive(False);
    neigh->SetAlive(True);
    neigh->E = current->E;
    Draw(cell_index);
    Draw(favourite);
    }

\end{verbatim}
\normalsize
Notice how we avoid referring to the specific details of the lattice
by only using the lists of neighbours which are returned by the
QueryNeighbours function.

As one would expect in a thermalizable system, equilibrium is
achievable here after a number of iterations.  The escape of a
particle from the lattice is possible if it can borrow enough thermal
energy from the local lattice site,
with a certain probability which may be defined in the
model. This results in a local cooling of the lattice.  The higher the
temperature of the lattice, the greater the probability of escape.  In
figure \ref{diff}c one sees lighter areas of locally higher
temperature. After a certain time one sees the equilibration in
relation to the environmental parameters in the model. The number of
particles in the lattice stabilizes: as the temperature increases and
evens out, the likelihood of particles escaping increases.

\section{Conclusions}

We have presented a framework for simultaneously visualizing and
computing numerical quantities from cellular or lattice simulations,
using a scheme of abstractions which faithfully separates independent
issues.  Our framework is based on an object oriented analysis of the
key elements of physical systems on a lattice.  As an example we
present the $XY$ model for electron spins at finite temperature and
show the approach to thermal equilibrium in both the vortex phase and
the ferromagnetic phase. Using the visual simulation it is possible to
see the Kosterlitz-Thouless phase transition in this model: tightly
bound pairs of vortices below a critical temperature, and vortex
unbinding above the critical temperature.  We are further able to
visualize the gauge invariant currents surrounding quantum mechanical
vortices in a magnetic field. A toy model for diffusion illustrates
the lattice independence possible in some cases.

Our computer programs are suitable for unix workstations running
release 6 of the X11 window system (Linux or FreeBSD for instance). More
information about how to collect and adapt these simulations may be
found at the WWW site http://www.iu.hioslo.no/\verb+~+cell. We may
also be contacted by email at case@iu.hioslo.no.

\subsection*{Acknowledgments}

We would like to thank F. Ravndal for stimulating this visualization
project and A. Sudb\o\ for introducing us to the interesting idea of
visualizing the $XY$ model in external magnetic field. We also very
much like to draw attention to the essential contributions made by
J. Mikkelsen and T.E. Sevaldrud in the design and implementation of
the CASE library as part of their dissertation in computer science at
Oslo College.

\section{Figure captions}

\begin{enumerate}

\item{The CASE object hierarchy.}\label{fig1}

\item  A $5\times 5$ lattice with spins. The sites ${\bf x}$ and ${\bf x'}$ are
            indicated. Here $\phi_{{\bf x}}- \phi_{{\bf x'}} = \alpha - \beta$.\label{fig2}

\item Vortices in the $XY$ model: (a) All spins point in the direction away from the 
vortex core. (b) The spins have all been rotated an angle $\pi/2$. This
whirl-shape is a characteristic of a vortex.
\label{Vortex}

\item This figure shows the route into equilibrium at $T=0.1$ for the $XY$
model. \label{XYmodel1}

\item Equilibrium snapshots at $T=0.8$ and $T=1.0$. The vortices are seen
to exist mostly in bound pairs. An example of vortex unbinding is shown
in b).\label{Video4}

\item The phase diagram of a type-II superconductor. \label{H-T}

\item The characteristic lengths of the model. While the coherence length 
$\xi$ equals the lattice spacing, the magnetic penetration depth
$\lambda$ is supposed to be much larger than the inter vortex
distance. \label{lambda}

\item When choosing a trial configuration a random phase $\Delta$ is added 
to the phases $\alpha$ of the four links connected to the probed site
in this manner. This corresponds to adding the phase $\Delta$ to the
superconducting phase $\phi$ at the site, which ensures that
Eq. ~\ref{LINKS} is always fulfilled and thus that the magnetic flux
through the four neighbouring plaquettes remains
unchanged. \label{MCSTEP}

\item Snapshots of the XY model in a magnetic field. (a) Shows an
initial disordered state of the system (b) Shows the state of the
system after a large number of iterations, where an almost hexagonal
lattice of the vortices is formed. (c) An enlarged portion of the
figure in (b) obtained by zooming in on the lower right hand corner
with the middle mouse button, a standard function in the CASE
library.\label{XYB}

\item Snapshots of the thermodiffusion simulations: a) shows the model
implemented on a square lattice, (b) shows the model on a triangular
lattice and c) illustrates the behaviour of the colour highlighting in
the triangular case after some time. \label{diff}

\end{enumerate}


\begin{thebibliography}{99}
\bibitem{footnote1} Note that our code is constructed in such as way as
to make this restriction removable at a later time.
\bibitem{footnote2} There are
several available tools for mathematical modeling: Mathematica,
Maple, IDL etc, but these are commercial programs designed for a much
wider purpose than we require, and are therefore inefficient for
numerical iteration combined with graphics. Also the new language Java
turns out to be too slow and not sufficiently mature to be useful for
our purposes, moreover they cannot handle the abstractions which
CASE uses for modelling a physical environment.
\end{thebibliography}
\end{document}